\documentclass[multphys,vecphys]{svmult}

\usepackage{makeidx}         
\usepackage{graphicx}        
\usepackage{multicol}        
\usepackage[bottom]{footmisc}

\makeindex             

\begin{document}

\title*{VIMOS total transmission profiles for broad-band filters}
\author{S. Mieske\inst{1}\and
M. Rejkuba\inst{1}\and S. Bagnulo\inst{2}\and C. Izzo\inst{1}\and G. Marconi\inst{2}}
{\institute{ESO, Karl-Schwarzschild-Str.2, 85748 Garching b. M\"unchen \and ESO, Alonso de Cordova 3107, Vitacura, Santiago}
}
%
%
\maketitle
\begin{abstract}
  VIMOS is a wide-field imager and spectrograph mounted on UT3 at the
  VLT, whose FOV consists of four 7'x8' quadrants.  Here we present
  the measurements of total transmission profiles -- i.e. the
  throughput of telescope + instrument -- for the broad band filters
  U, B, V, R, I, and z for each of its four quadrants. Those
  measurements can also be downloaded from the public VIMOS web-page.
  The transmission profiles are compared with previous estimates
  from the VIMOS consortium.
\end{abstract}

\section{Transmission profiles of telescope + instrument}
\begin{figure}[]
\centering
\resizebox{\hsize}{!}{
\includegraphics[width=7.3cm,height=6.5cm]{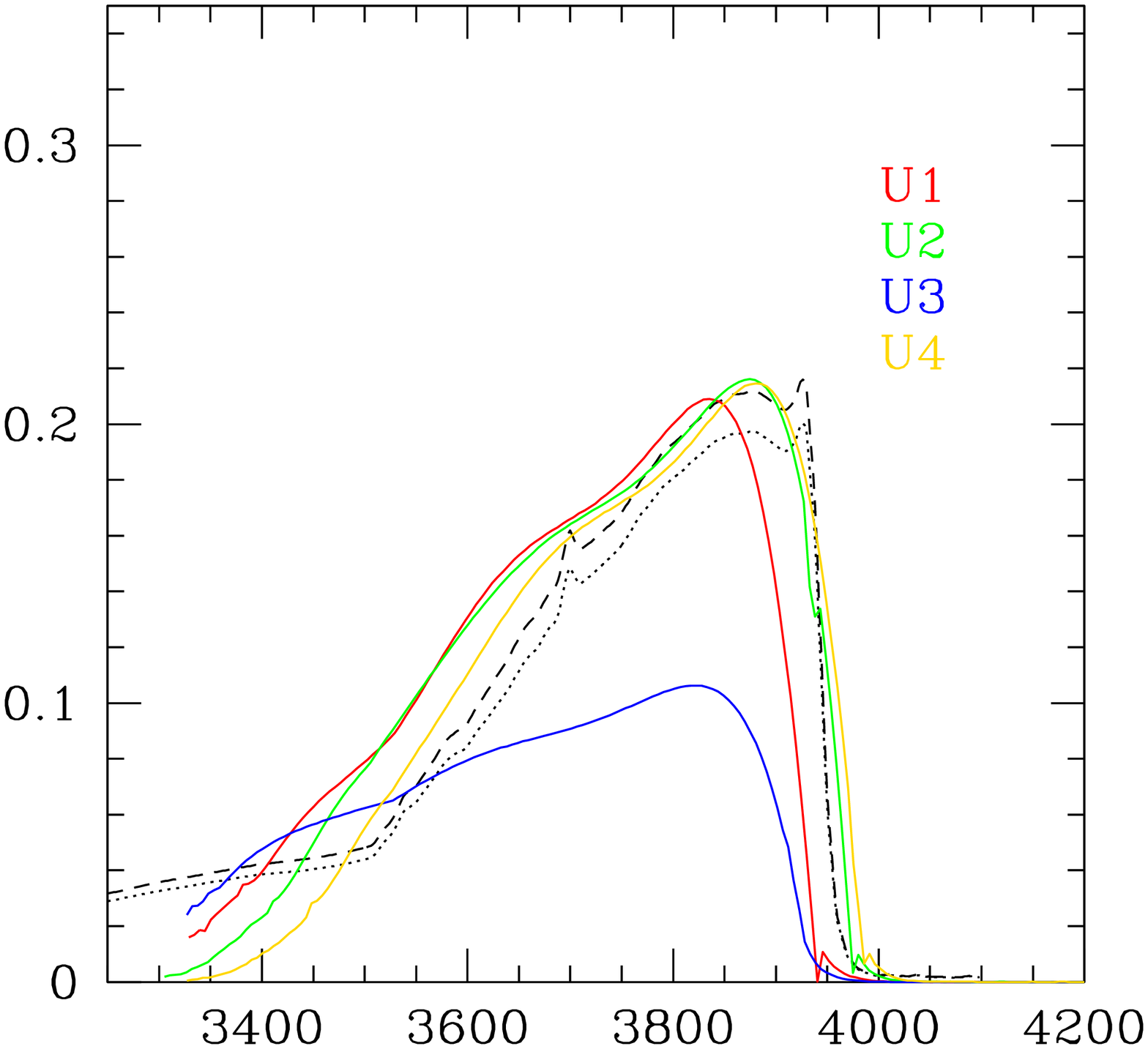}
\includegraphics[width=7.3cm,height=6.5cm]{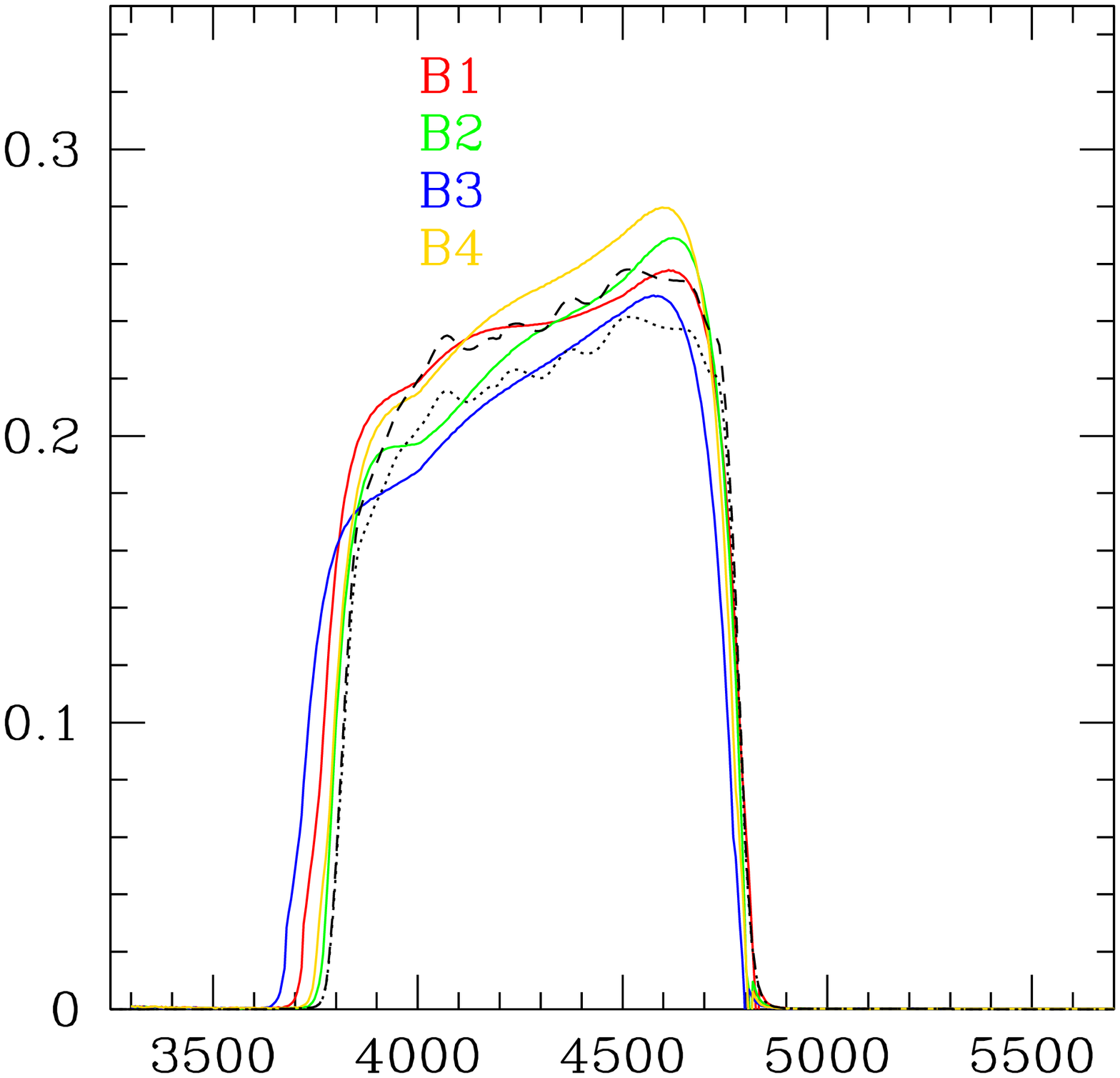}}
\resizebox{\hsize}{!}{
\includegraphics[width=7.3cm,height=6.5cm]{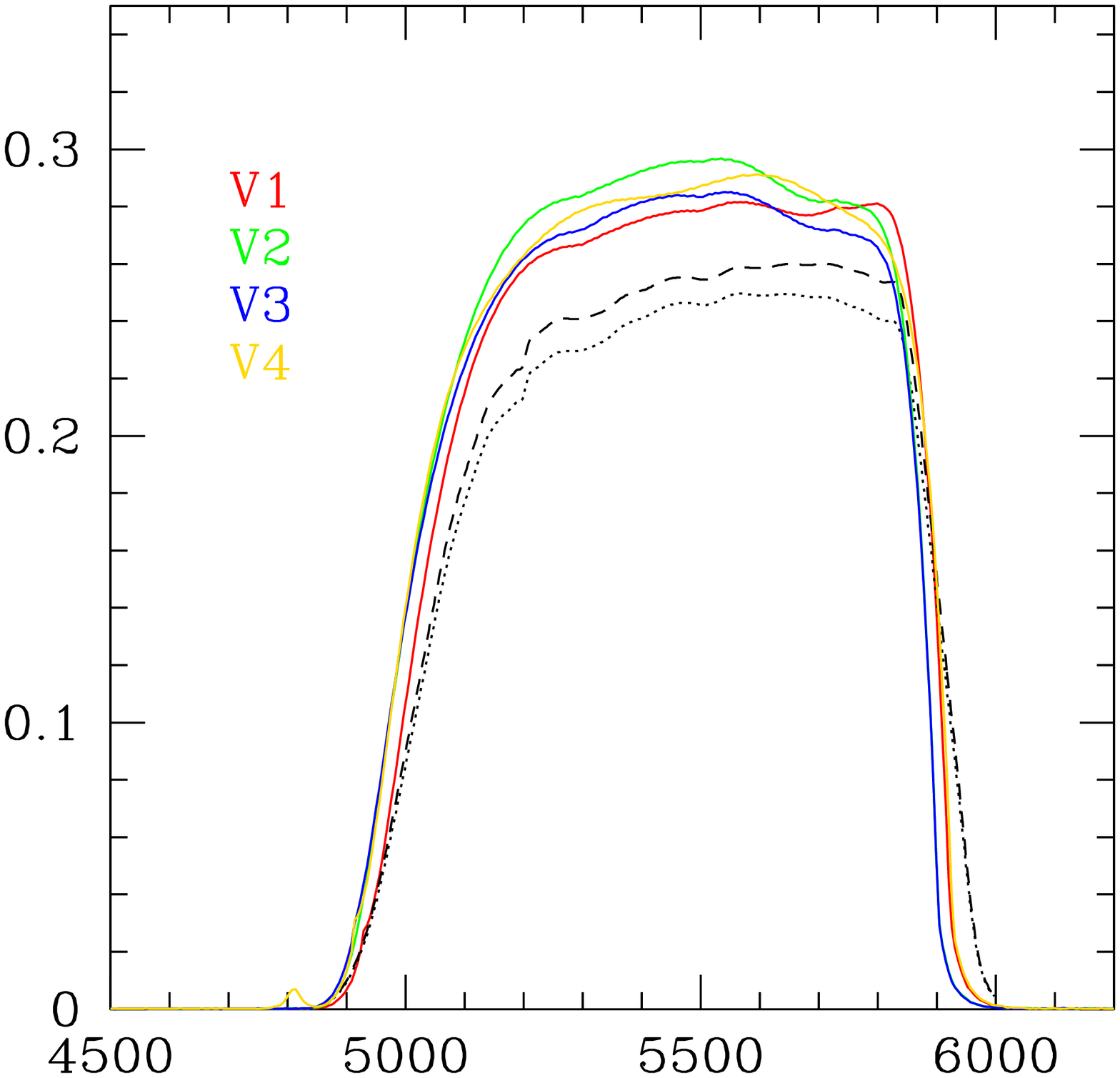}
\includegraphics[width=7.3cm,height=6.5cm]{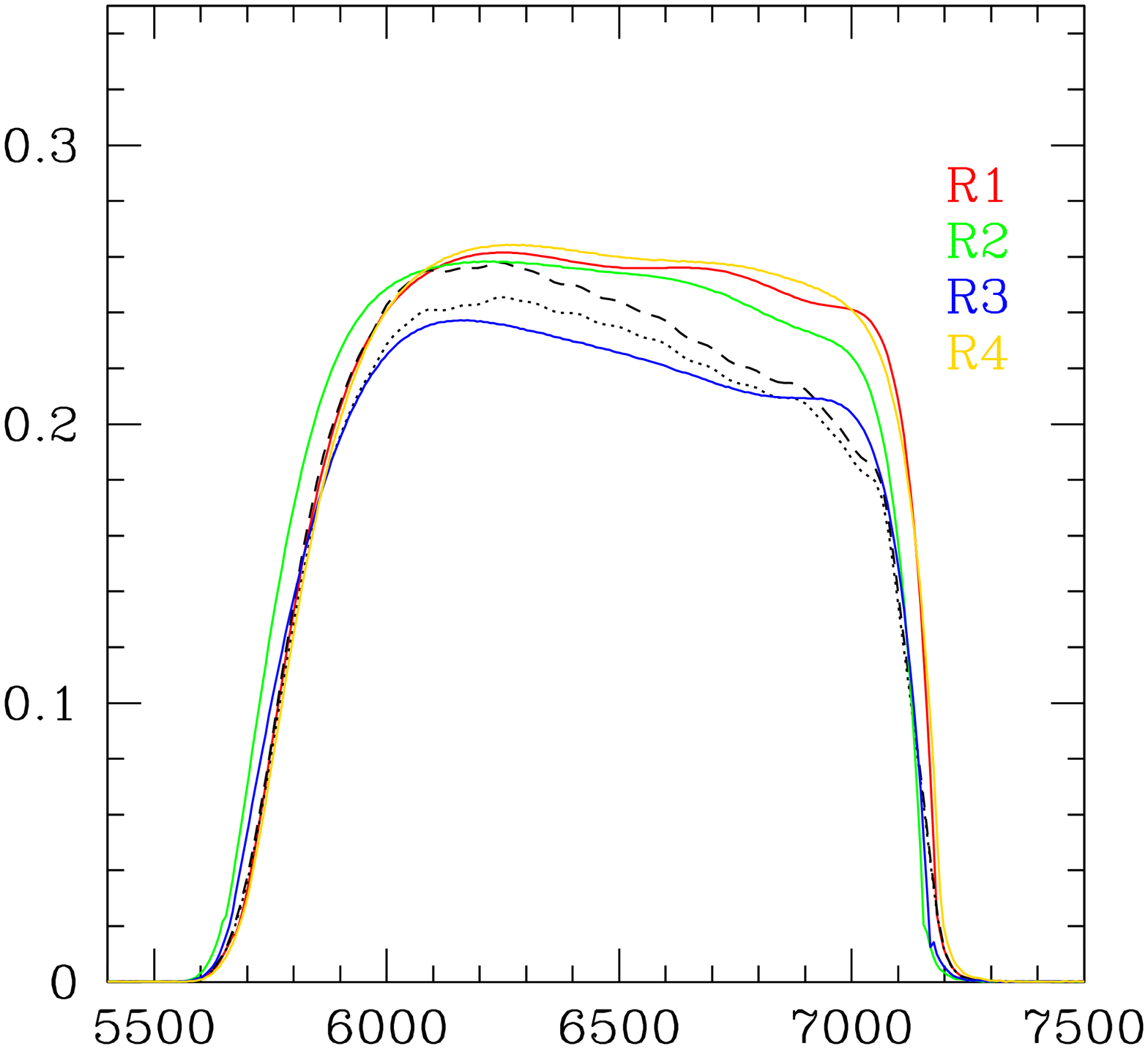}}
\resizebox{\hsize}{!}{
\includegraphics[width=7.3cm,height=6.5cm]{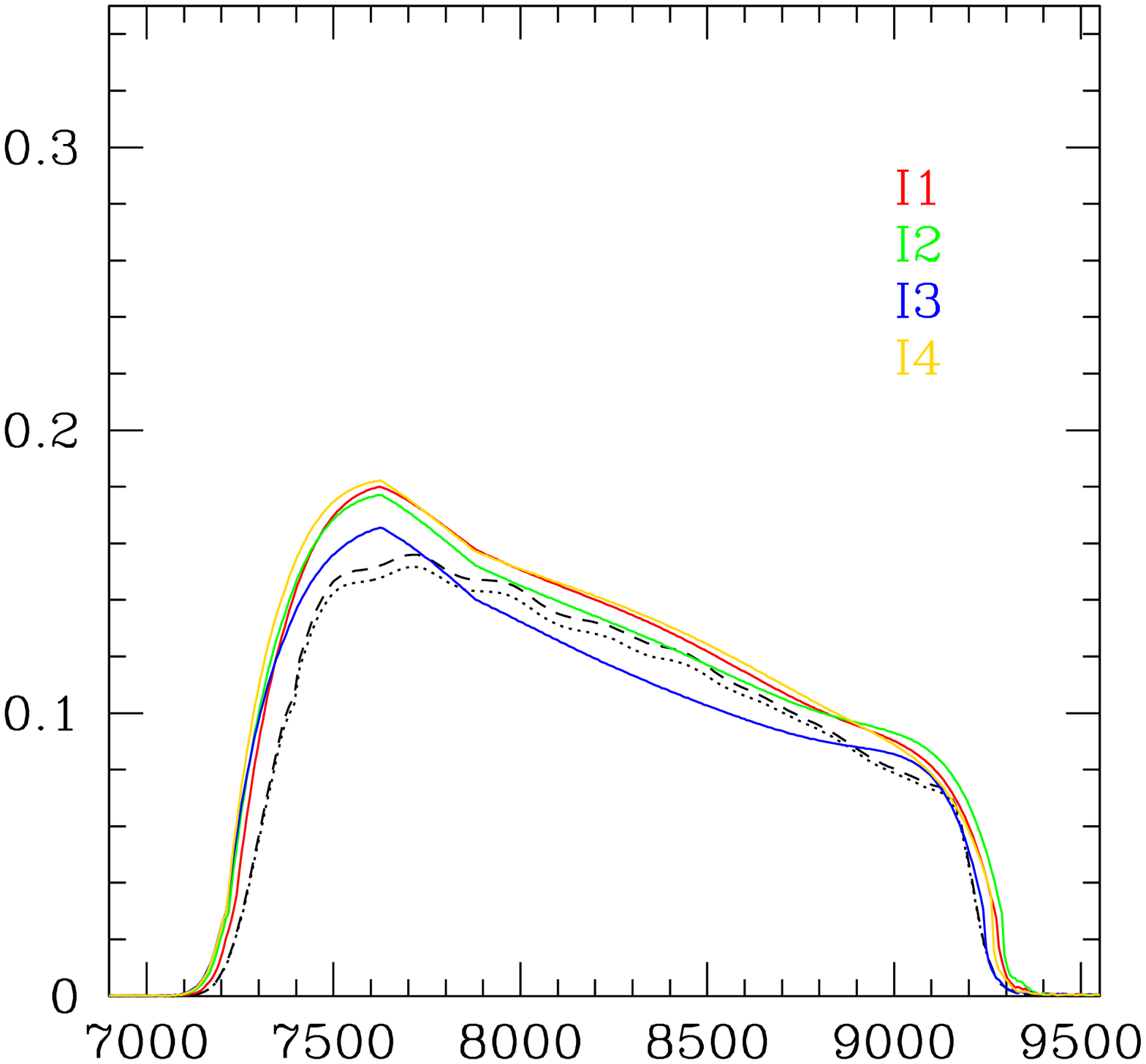}
\includegraphics[width=7.3cm,height=6.5cm]{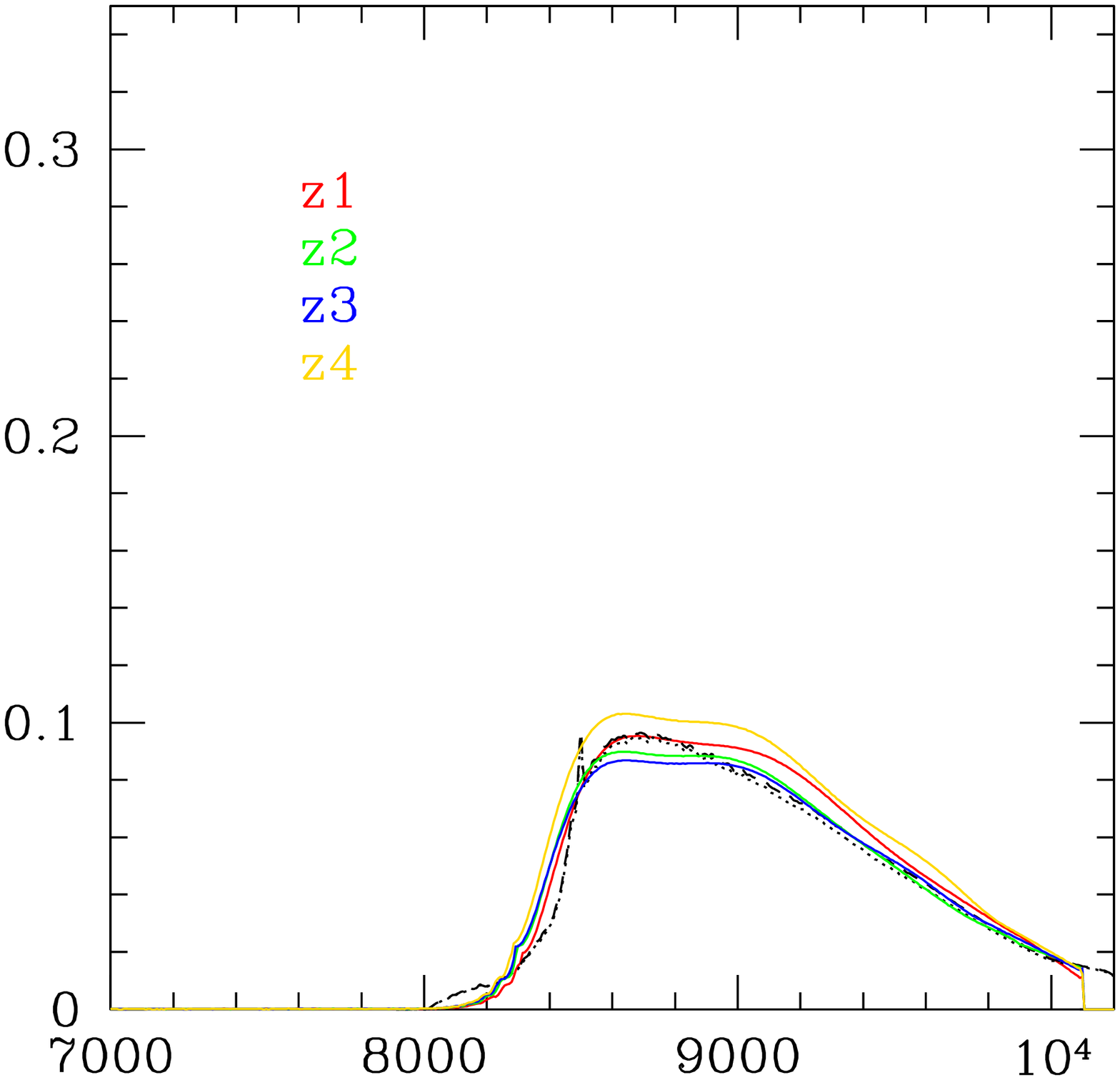}
}
\caption{\label{smieske:fig1}. Total VIMOS telescope+instrument efficiency curves for the broad band filters U, B, V, R, I, and z. X-axis is in units of \AA ngstrom.  The dotted and dashed lines are the efficiency curves available from the VIMOS Exposure Time Calculator  on the web, assuming the La Silla (dotted) and CTIO (dashed)  extinction curve.  The U-band estimate is based on the star LTT 7379. The B-band estimate is the mean of the efficiency derived from Hiltner 600 and LTT 7379.  The VRIz estimates are based on observations of Hiltner 600.}
\end{figure}

What has been available to ESO up to now are efficiency curves
provided by the VIMOS consortium that are averaged over the four
quadrants and were determined with laboratory measurements, not by
observing standard stars.  Prompted by user requests to have separate
efficiency estimates for each quadrant, two spectrophotometric
standard stars were observed in 2006: Hiltner 600 for all filters, and
LTT7379 for the U and B band. These observations were performed with
the broadest possible slit (5$"$) and the grisms LR-blue (for U, B,
and V) and LR-red (for R, I, and z).  Under the assumption that the
total star-light is contained in the slit, the efficiency curve is
derived from the measured flux. The required input consists of the
tabulated standard star flux, the extinction curve, and the grism
transmission as provided by the VIMOS consortium. The resulting curves
are shown in Fig.~\ref{smieske:fig1}. They are also available from the
public VIMOS
web-pages\footnote{http://www.eso.org/instruments/vimos/inst/imaging.html}.

There is good agreement between the newly derived efficiency estimates
and those from the consortium.  The substantial sensitivity drop in
quadrant 3 for the U-band is found independently for both LTT 7379 and
Hiltner 600 (note that the measurement for Hiltner 600 was not used
for the efficiency estimate in the U-band since it suffered from
flux loss in all four quadrants, due to a slight misalignment
of the standard within the slit). Since the U-band photometric
zero-points for Q3 do not show such a strong drop, a centring problem
due to instrument flexure is a possible reason for this lack of
measured flux. In addition, the consortium estimate of the grism
transmission in Q3 for l$<$3800 {\AA} may be too high. We suggest that
the efficiency estimates especially in the U-band be always re-scaled
to the (integrated) photometric zero-points.

\section{Transmission profiles of broad-band filters}
\begin{figure}
  \centering \resizebox{\hsize}{!}{
    \includegraphics[width=8cm]{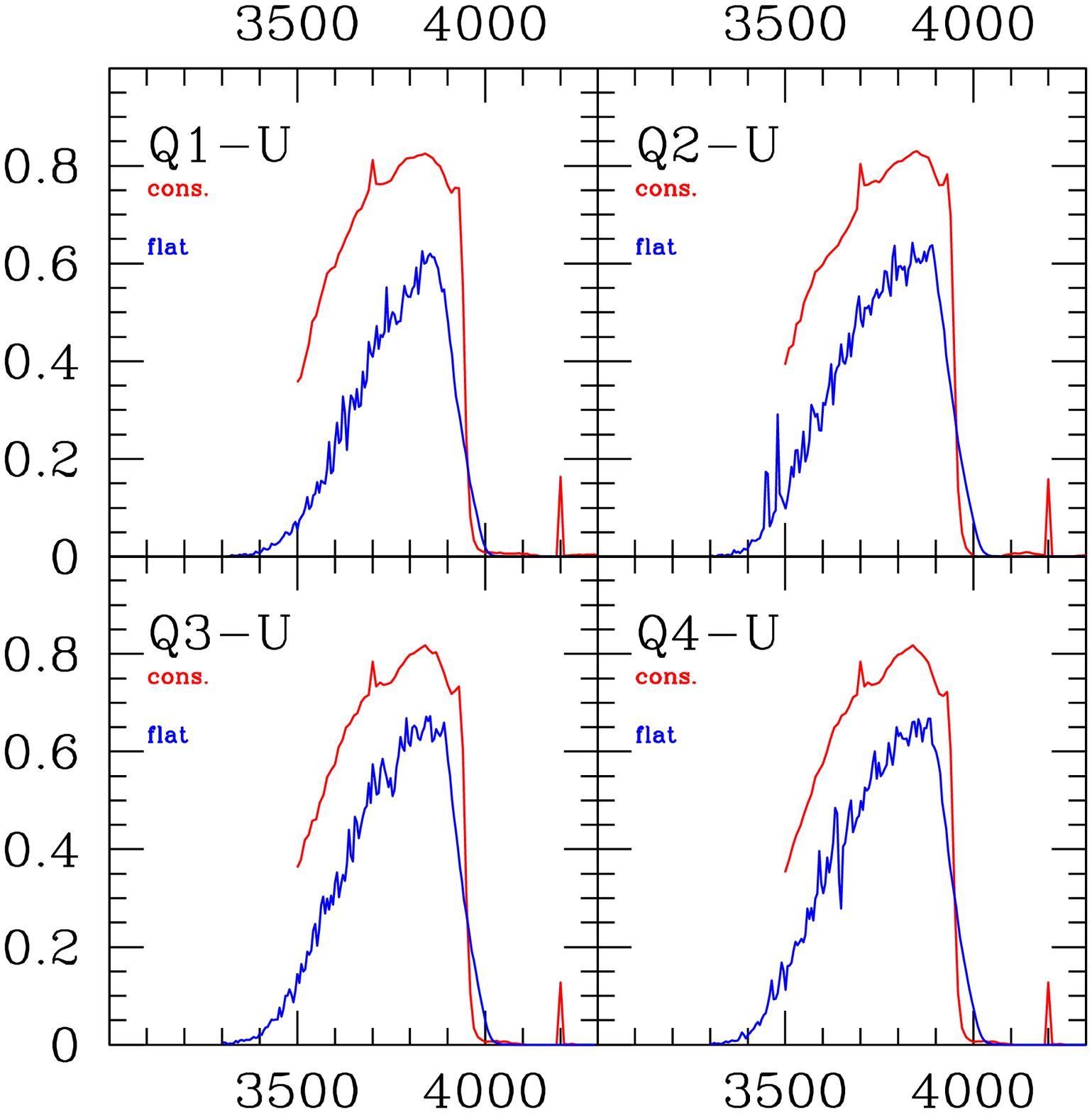}
    \includegraphics[width=8cm]{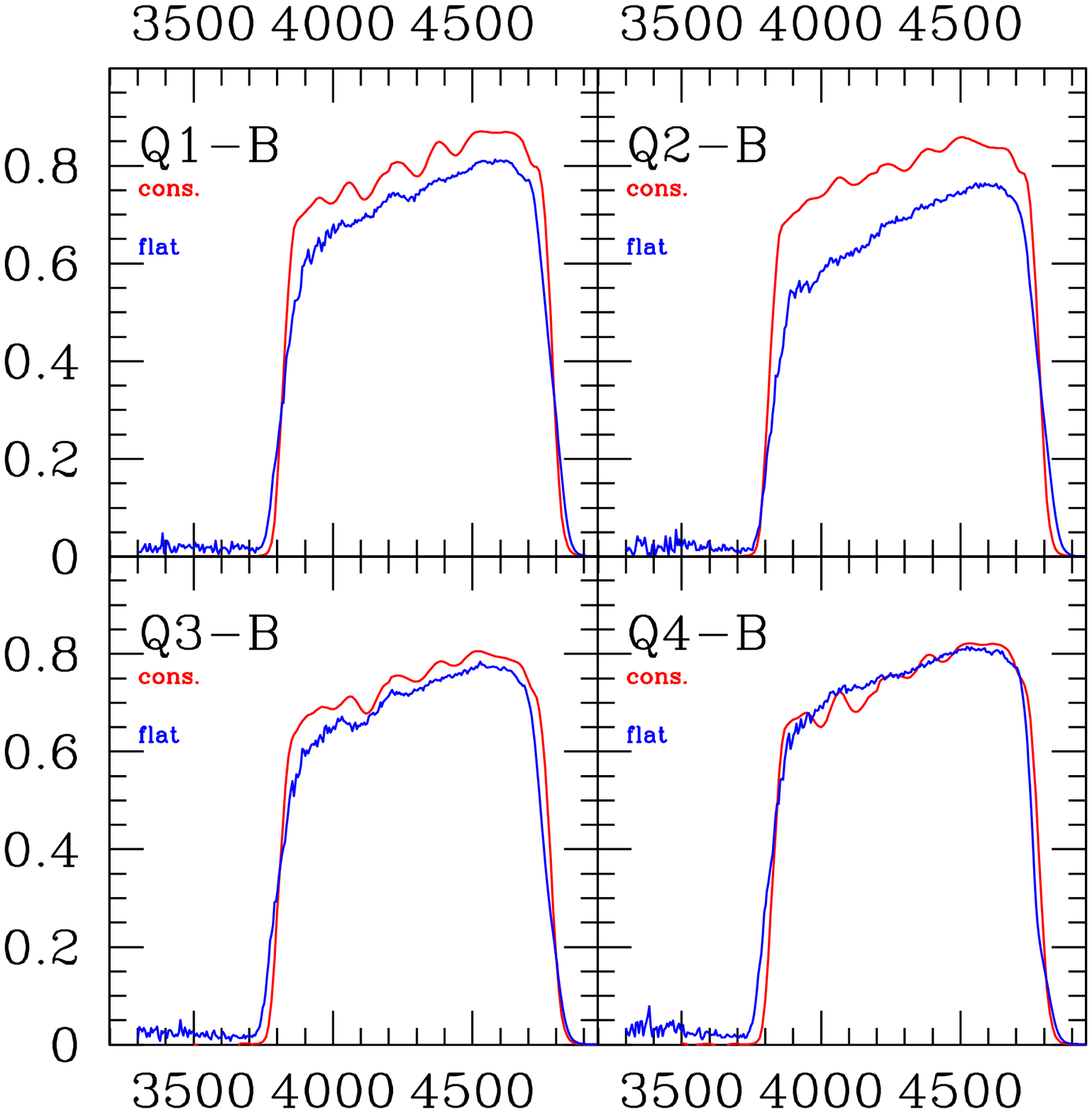}}
  \resizebox{\hsize}{!}{
    \includegraphics[width=8cm]{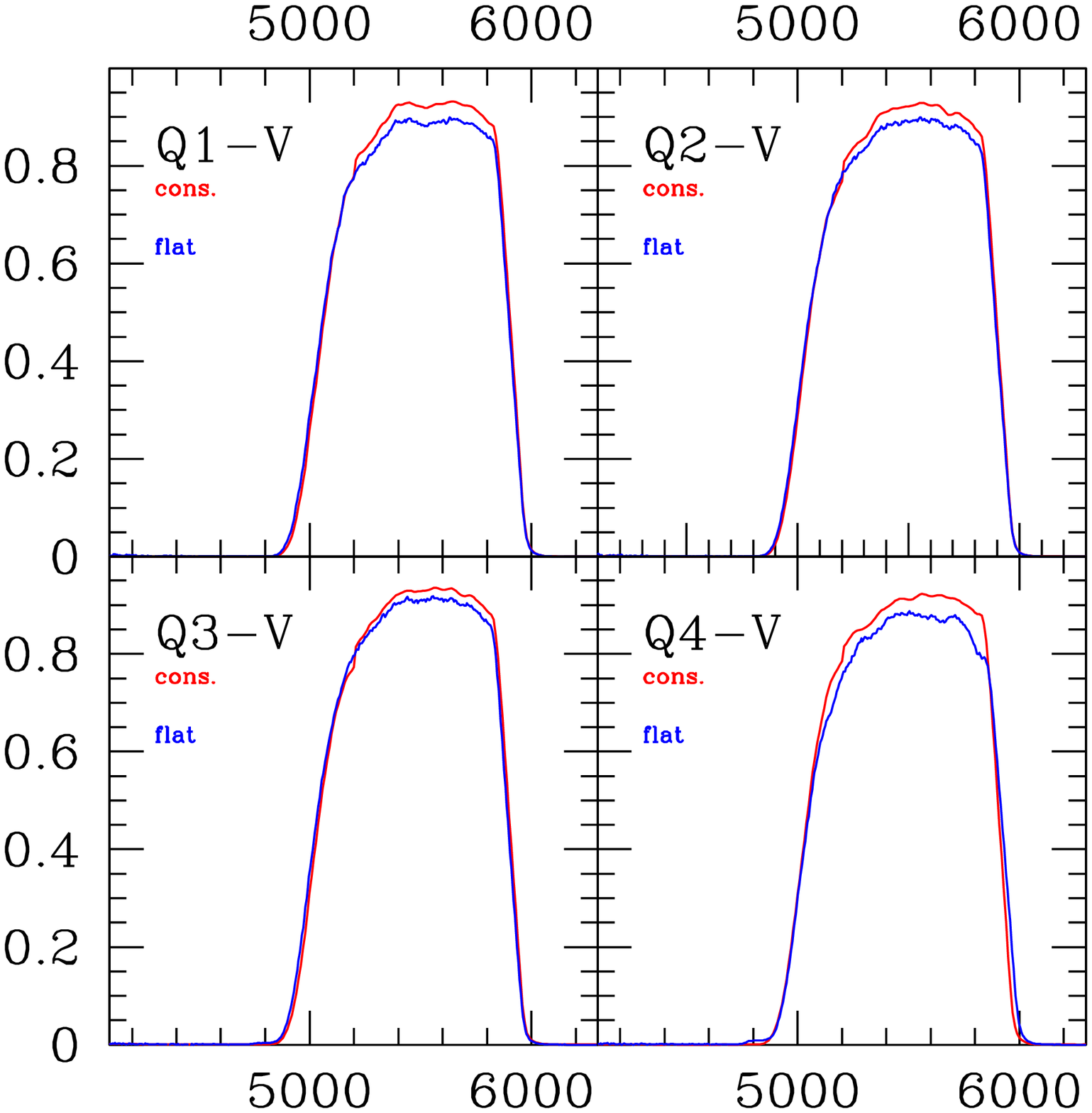}
    \includegraphics[width=8cm]{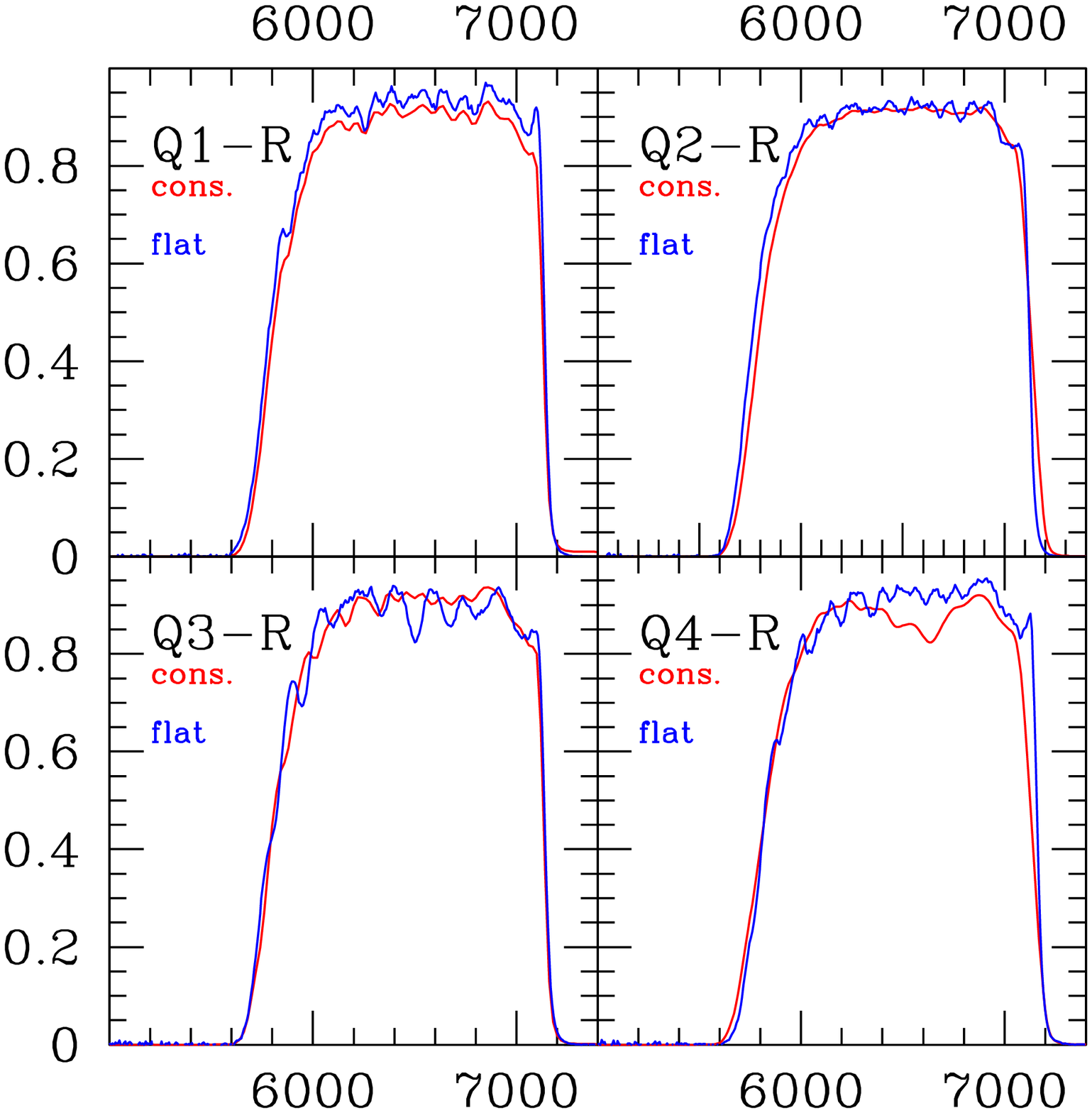}}
  \resizebox{\hsize}{!}{
    \includegraphics[width=8cm]{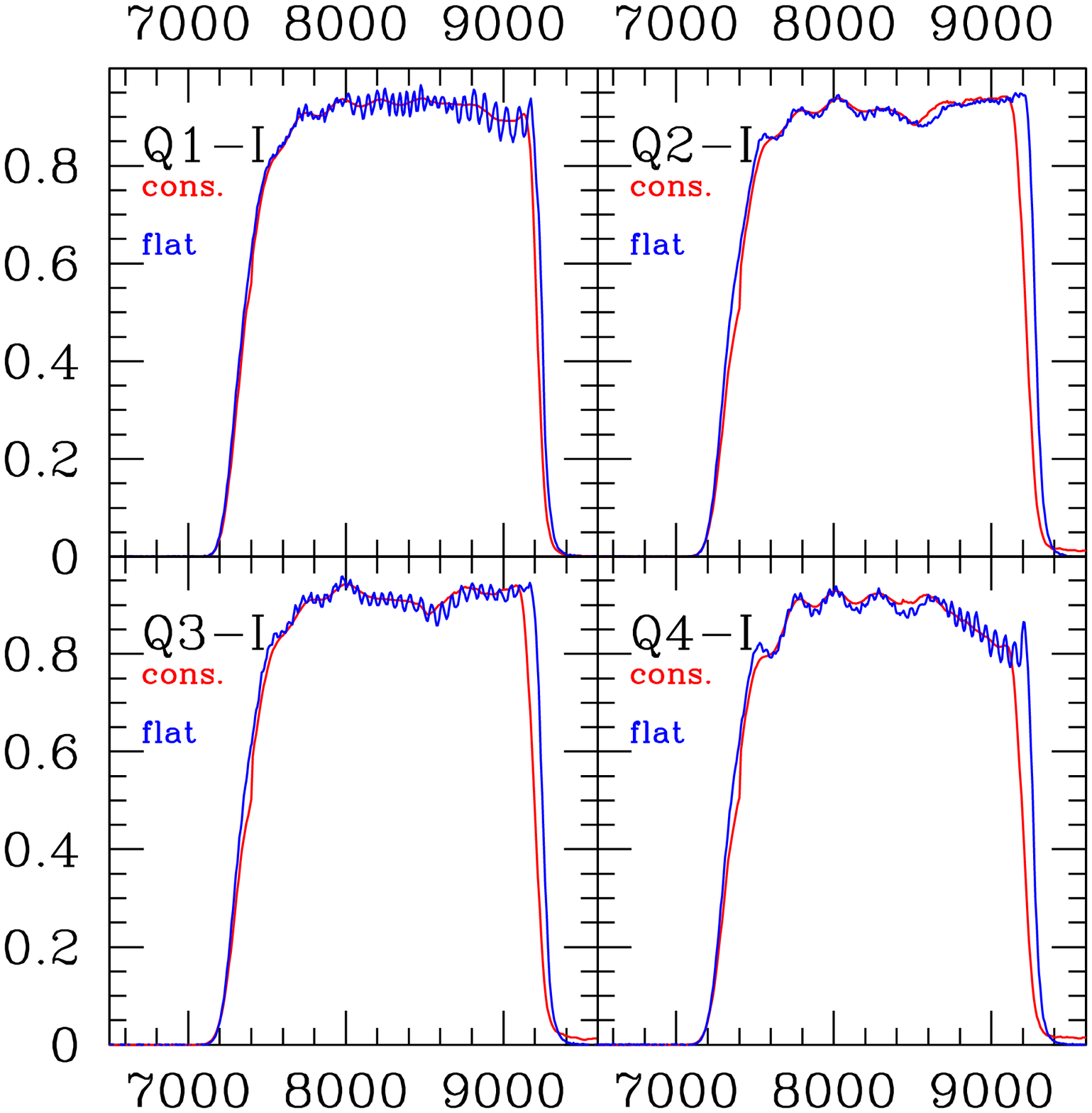}
    \includegraphics[width=8cm]{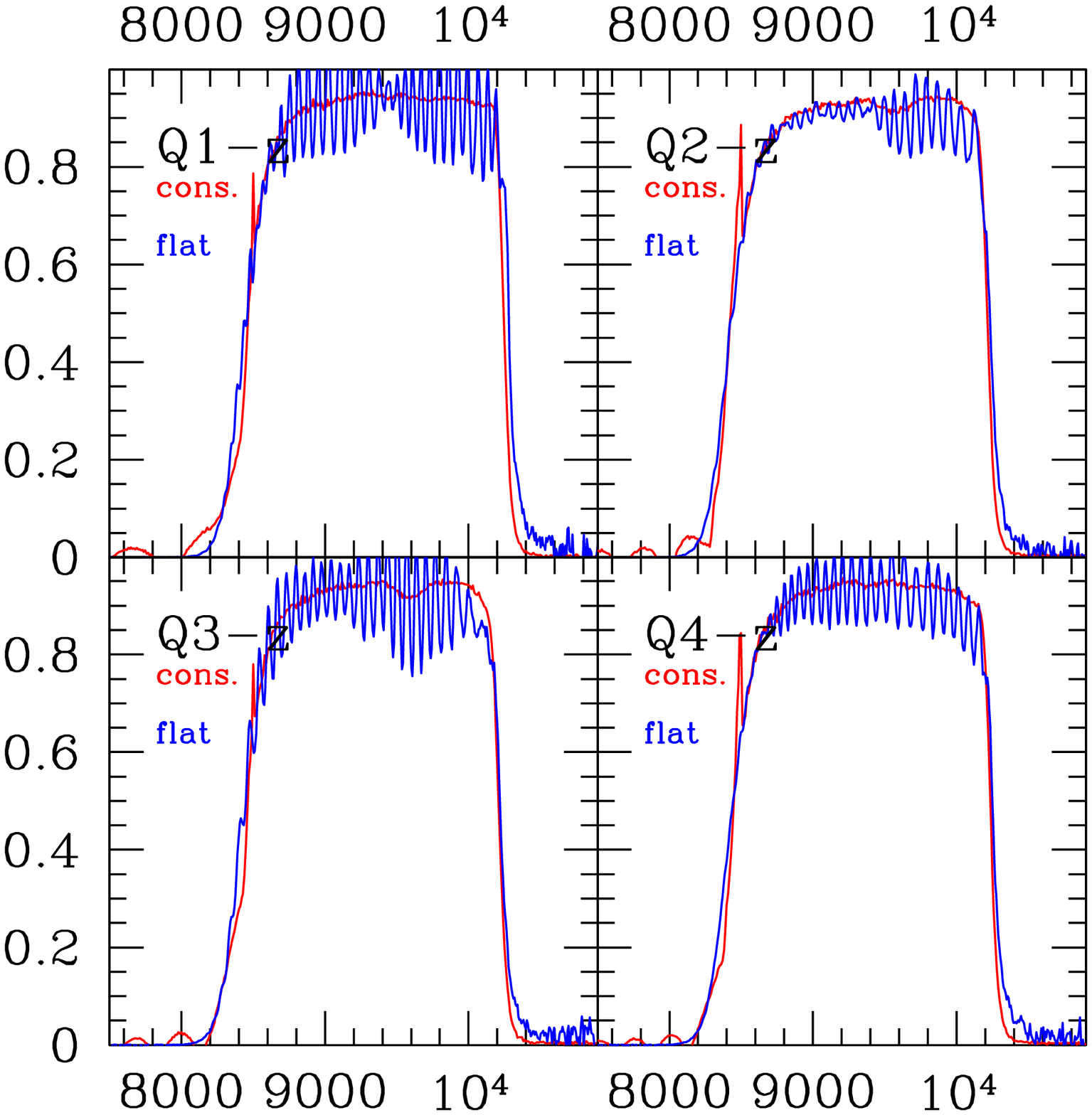} }
\caption{\label{smieske:fig3}. Blue lines indicate filter transmission profiles for the VIMOS broad band filters as derived from continuum lamp exposures taken with and without the filters inserted. Red lines indicate
the  transmission estimates provided by the VIMOS consortium.}
\end{figure}

In addition to the total instrument+telescope efficiency estimates, we
used the continuum lamp screen flats taken with and without inserted
filter to measure the filter transmissions for each quadrant. The
results are shown in Fig.~\ref{smieske:fig3} and compared with the
consortium estimates. There is very good agreement for the V,R,I, and
z filters. For the U-band, the consortium estimates are about 20\%
above those derived from the screen flats. Also for the B-band, the
consortium estimates are about 5-10\% higher.

\begin{figure}[]
\centering
{
\includegraphics[width=8cm,height=8cm]{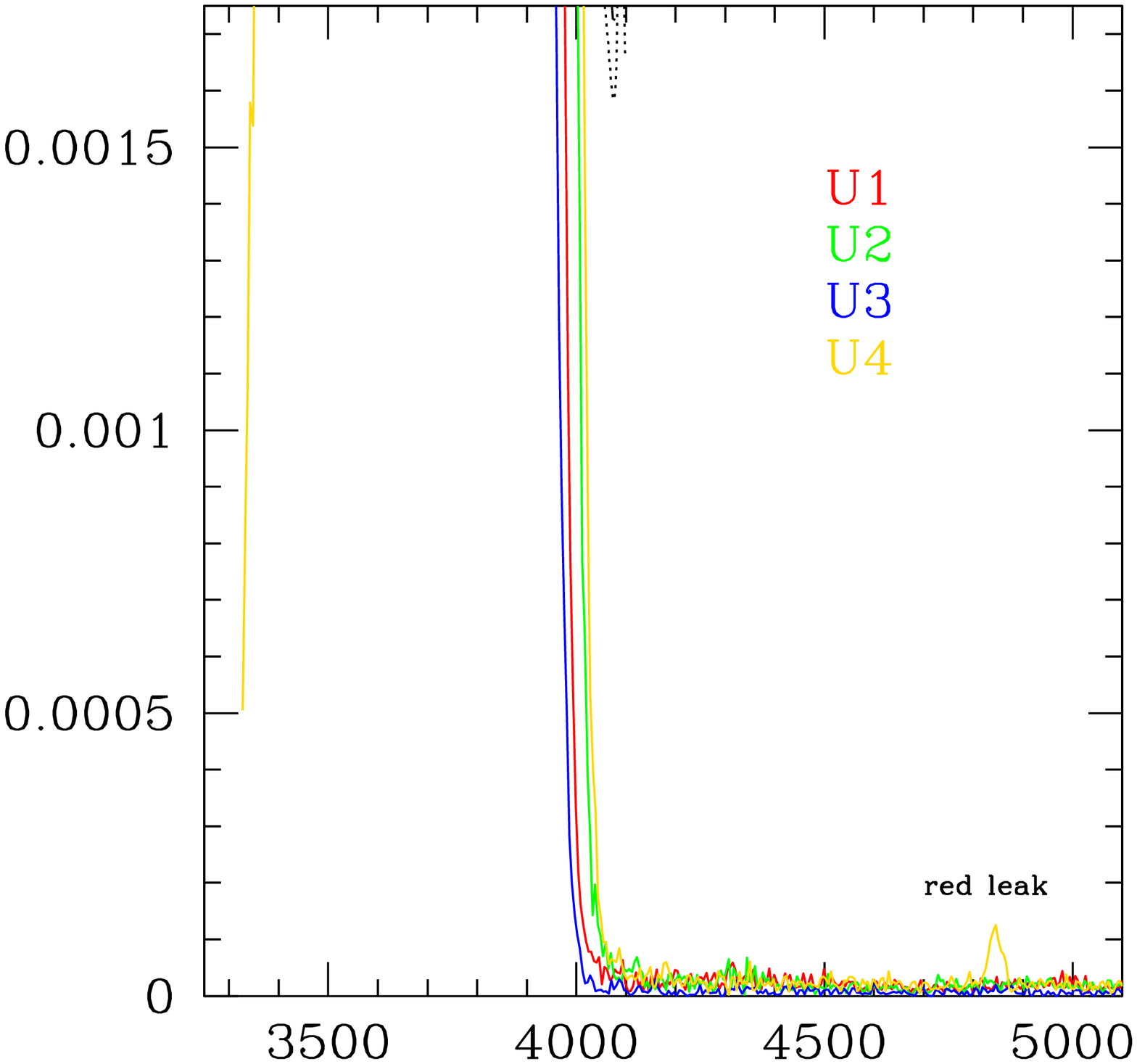}
}
\caption{\label{smieske:fig2} Detection of a very small ``red leak'' in the U-band efficiency curve of Q4. Note that the y-axis has been zoomed in by a factor of 200 compared to Fig.~\ref{smieske:fig1}. The peak of the red leak is about 0.1\% of the efficiency peak.}
\end{figure}

Unlike in the consortium filter transmission curves from
Fig.~\ref{smieske:fig3}, we do {\it not} detect a red leak of
the U-band filter around 4200 \AA. Only in Q4, there is a very
minor leak at about 4850 {\AA} (see Fig.~\ref{smieske:fig2}).

\section{Conclusions}
The measured telescope+instrument efficiency curves for the VIMOS
broad band filters agree very well with the consortium estimates. A
discrepancy in the U-band is observed for one quadrant, possibly due
to flexure and low grism transmission at short wavelengthes.  The
filter transmissions also agree very well, except for the B and
especially U-band, where the newly measured transmission is about 20\%
lower.
%
%


\printindex
\end{document}